\documentclass[12pt]{article}
\usepackage{amsmath}
\usepackage{epsfig}
\usepackage{epstopdf}
\usepackage{amssymb}

\newcommand{\beq} {\begin{equation}}
\newcommand{\eeq} {\end{equation}}

\begin{document}

\begin{flushright}
HIP-2010-36/TH
\end{flushright}

\vskip 2cm \centerline{\Large
{\bf Solitons as Probes}}
\vskip 0.3cm
\centerline{\Large {\bf of the Structure of Holographic Superfluids}}
\vskip 1cm
\renewcommand{\thefootnote}{\fnsymbol{footnote}}
\centerline
{{\bf Ville Ker\"anen,$^{1,2}$
\footnote{ville.keranen@helsinki.fi}
Esko Keski-Vakkuri,$^{1,2}$\footnote{esko.keski-vakkuri@helsinki.fi}
Sean Nowling,$^{3}$\footnote{nowling@nordita.org}
K. P. Yogendran,$^{4}$ \footnote{yogendran@iisermohali.ac.in}
}}
\vskip .5cm
\centerline{\it
${}^{1}$Helsinki Institute of Physics }
\centerline{\it P.O.Box 64, FIN-00014 University of
Helsinki, Finland}
\centerline{\it ${}^{2}$Department of Physics}
\centerline{\it P.O.Box 64, FIN-00014 University of Helsinki, Finland}
\centerline{\it
${}^{3}$NORDITA}
\centerline{\it Roslagstullsbacken 23, SE-106 91 Stockholm, Sweden}
\centerline{\it
${}^{4}$IISER Mohali}
\centerline{\it MGSIPAP Complex, Sector 26
   Chandigarh 160019 , India}
\setcounter{footnote}{0}
\renewcommand{\thefootnote}{\arabic{footnote}}

\begin{abstract}
Detailed features of solitons in holographic superfluids are
discussed.
Using solitons as probes, we study the behavior of holographic superfluids by
varying the scaling dimension of the condensing operator, and build a comparison
to the BEC-BCS comparison phenomena.
 Further evidence for this analogy is provided
by the behavior of solitons' length scales as well as by the
superfluid critical velocity.
\end{abstract}

\newpage

\section{Introduction}
In recent years there has been great interest in studying superfluidity
in ultracold Fermi gases.  In the laboratory, one cools a gas of
fermionic atoms to temperatures below a few microKelvins. By
subjecting the system to a controllable external magnetic field, one
can tune the interactions in the fermionic gas. Famously, what one
finds is a crossover phenomenon.  On one side of the range the system
is described by condensed loosely bound Cooper pairs of fermions;
while at the other end the fermions become very strongly bound and the
system is effectively characterized as a condensate of a fundamental
bosonic degree of freedom. These two extremes are separated by the
unitarity point, where the systems do not have a simple description in
terms of either simple bosons or fermions.

At zero temperature, precisely at the unitarity point, apart from the
fermion density the system has no other length scale.  This feature is of
interest to holographic model building in string theory, where it has
been discovered that some strongly coupled scale invariant theories
have a dual description in terms of gravitational theories in
asymptotically anti-de Sitter spacetimes of one higher dimension.

One of the first holographic models for a superfluid was introduced in
\cite{Hartnoll:2008vx}\footnote{The model was actually presented as a
  holographic model for some aspects of superconductivity, but a more
  precise interpretation is that of a superfluid.}. This is a
relativistic $2+1$ dimensional superfluid described holographically
using a gravitational theory in $3+1$ dimensions.  Using classical
gravity theory, one expects to reproduce a large $N$ strong coupling
limit of a dual $2+1$ dimensional system, which we will often refer to
as the (dual) field theory.  Typically one expects the field theory to
possess both bosonic and fermionic degrees of freedom, and the
operator responsible for symmetry breaking could then be a composite
operator including fermions (see {\em e.g.} the discussion in
\cite{Gubser:2009qm}).  It was shown in \cite{Hartnoll:2008vx}, that
for low enough temperatures, the $U(1)$ global symmetry of the field
theory is broken and that this phase, in the dual gravity description,
presents itself as a black hole with complex scalar hair outside its
horizon.

It is important to consider to what extent the model
\cite{Hartnoll:2008vx} can mimic the salient features of cold atomic
systems.  It is obvious that it cannot be literally interpreted as a
model for such systems.  The cold atoms are a $d=3$ non-relativistic
system, while the holographic system describes a relativistic $d=2$
large $N$ (mean field\footnote{We use the phrase ``mean field'' in the
  sense of expanding around a ground state - which is not necessarily
  the gaussian one. The classical gravity solution, provided it is
  stable, defines a saddle point for a perturbative expansion which
  maps into a perturbation expansion on the field theory side. While
  the elements of expansion are not identical across the two
  descriptions, they are nevertheless corresponding.})
theory. Furthermore, it would be natural to explore the possibility to
find a gravity dual for a system at unitarity in view of the conformal
invariance of the gravity description. We will address the question
whether there is any sense in which \cite{Hartnoll:2008vx} could
resemble such a system?

Firstly, it is encouraging to note that many of the features of the
non-relativistic condensates are expected to have relativistic
analogs, since the basic intuitions about the nature of bound states
are the same \cite{Nishida:2005ds}.  Secondly, the unitarity regime
was also studied without introducing a dimensionful parameter, using
an $\epsilon$ expansion \cite{Nishida:2006a,Nishida:2006b}.

Further \cite{Nishida:2006a,Nishida:2006b} one way to interpret the
$\epsilon$ expansion study of the unitarity regime is as a family of
conformal field theories separating a BCS like unitarity phase (near
$d=2$) from a BEC like unitarity phase (near $d=4$) with the most
interesting and strongly coupled unitarity near $d=3$.  There is a
natural parameter controlling the physics of the holographic
condensates, namely the scaling dimension of the condensing
operator. Changing the scaling dimension might be interpreted as a
family of conformal field theories corresponding to different kinds of
superfluids on the unitary regime as in
\cite{Nishida:2006a,Nishida:2006b}
\footnote{The analogy between our family of fixed points to the one in
  \cite{Nishida:2006a,Nishida:2006b} cannot be taken too literally
  since in \cite{Nishida:2006a,Nishida:2006b}, (see also
  \cite{Sachdev}) the condensate scaling dimension is held fixed while
  $\epsilon$ is varied.}. Finally we would like to point out that
\cite{Nishida:2005ds} found that, at least when approaching unitarity
from the BEC side, relativistic and non-relativistic systems behave
similarly.

The fact that this system is strongly coupled, involves both fermion
and boson degrees of freedom, and may be studied using conformal field
theory methods, suggests that it may have a degree of commonality with
holographic theories \cite{Nishida:2007pj,Son:2008ye,Balasubramanian:2008dm}.
To summarize, we adopt the point of view that the unitarity regime of
cold atoms provides a nice guide to study, interpret and organize many of the
features of the superfluid first described in \cite{Hartnoll:2008vx}, and
likely other related models as well.

What would then be a suitable probe to the subtle microscopic features
at both sides of unitarity? It is interesting to note that solitons in
a superfluid phase provide nice probes of the short distance structure
even at the mean field level, \cite{antezzaetal,Randerias} (at least
away from the strict unitarity limit, where constructing the mean
field theory is problematic).  The main reason for this is that across
the soliton's core the superfluid must interpolate all the way from
the broken to the symmetry restored phase.  The core's structure will
have features which may be used to characterize the short distance
features of the theory\footnote{To our knowledge, such an exploration
  of solitonic features has not been conducted in the context of an
  $\epsilon$ expansion approach.}. An interesting example of the use
of solitons in understanding the microscopic constituents of a certain
supersymmetric quantum field theories may be found in
\cite{Collie:2009iz}.

As will be described, solitons in the holographic system possess
several similarities with solitons found in cold atomic systems.  In
addition we will find that the analogy with atomic systems is useful
to organize certain linear response calculations in the holographic
system. Conversely, the properties of holographic solitons may be
of interest for the study of their real world counterparts. Holographic duals
provide a rather simple effective theory for a strongly coupled system at finite temperature.
In the case of superfluids, they give a toy model for the condensate and the normal component
and their interactions. For example, we can study how the charge depletion at the core of the soliton
varies as a function of the temperature, which would be very difficult to do using standard
condensed matter techniques.


\section{One holographic model of superfluidity\label{sec:Background}}

Ref. \cite{Hartnoll:2008vx} constructed a holographic dual for a
superfluid in 2+1 dimensions, provided by an Einstein-Abelian-Higgs
system in 3+1 dimensional asymptotically $AdS$ space,
\beq \label{eq:Action}
S_{AdS}=\int d^4x\sqrt{-g}\Big[\frac{1}{\kappa_4^2}\left(\mathcal{R}
+\frac{6}{L^2}\right)-\frac{1}{q^2}\Big(|D_{\mu}\Psi|^2
+m^2|\Psi|^2+\frac{1}{4}F_{\mu\nu}^2\Big)\Big]
\eeq
By virtue of the AdS/CFT correspondence, the gauged symmetries of the
gravitational (bulk) theory correspond to global symmetries of the
dual field theory (which shall also be called the ``boundary
theory''). Thus, the Lorentz covariance of the boundary theory follows
from the bulk diffeomorphism invariance and the bulk U(1) gauge
symmetry gives rise to a U(1) global symmetry in the dual field
theory. We shall identify the corresponding conserved charge with the
particle number in the language appropriate for atomic systems.

Spontaneous breakdown of this global symmetry will then result in a
charged superfluid condensate. The formation of the condensate is
dual, in the bulk gravitational theory, to spontaneous gauge symmetry
breaking - that is to say, to the condensation of a scalar field
charged under the gauge symmetry. The entire system maybe placed at
finite temperature by including a black hole background, in which case
the Hawking temperature corresponds to the equilibrium temperature of
the superfluid.

Basic properties of superfluids, such as transport properties at
linearized level
\cite{Herzog:2008he,Basu:2008st,Herzog:2009md,Yarom:2009uq,Amado:2009ts}
have been studied using the action (\ref{eq:Action}); the derivation
of non-linear superfluid dynamics has been discussed in
\cite{Sonner:2010yx}.

The AdS-CFT dictionary tells us that the gravitational action in $4$
dimensions evaluated on the solutions to the equations of motion (with
suitable boundary conditions) is the generating functional of various
moments of the corresponding operators in the $2+1$ dimensional quantum
field theory
\beq
e^{-S_{AdS}(on shell)}=\langle e^{-\int d^{3}x (\rho(x)\mu(x)+\mathcal{O} (x)J(x))}\rangle_{QFT} \ .
\eeq
The metric of the bulk spacetime is required to asymptote to Anti de
Sitter space (the asymptotic behaviour in fact defines the energy
momentum tensor of the field theory).
The boundary QFT operators are related to the
behavior of the bulk fields in the asymptotically AdS regime by the
relations
\beq\label{eq:SeriesExp}
\Psi(x,z)=z^\Delta \langle \mathcal{O}_{\Delta}(x)\rangle+z^{3-\Delta} J(x)+...
\quad A_{t}(x,z)=\mu(x)+z\rho(x)+...
\eeq
where the mass of the bulk field $\Psi$ is
\beq\label{eq:MassDelta}
m^2 = \Delta(\Delta-3).
\eeq
In the dual field theory one identifies $J$ with the source for the
operator $\mathcal{O}_{\Delta}$ of scaling dimension $\Delta$ in the dual field
theory.  The value $\langle\mathcal{O}_{\Delta}(x)\rangle$ occurring in
(\ref{eq:SeriesExp}) is identified with the expectation value of the
dual operator given an external source, $J(x)$. Also $\mu(x)$ is
interpreted as a chemical potential in the dual field theory, while
$\rho(x)$ is the (canonically) conjugate variable, the
charge(/particle number) density.

To have a well defined solution one must impose boundary conditions
which fix the value of $J(x)$ in the gravitational theory.  The value
of $\mathcal{O}_{\Delta}(x)$ is then determined uniquely.  In the
remainder we will choose our units so that the AdS scale is fixed to
$L=1$. For large $m^2$ one only finds a solution if $\Delta$ is chosen
to be the largest root of (\ref{eq:MassDelta}).  However,
for\footnote{We remind the reader that masses in AdS spacetimes are
  permitted to be slightly negative ($m^2>-9/4$) without triggering an
  instability so long as they are above the Breitenlohner-Freedman
  (B.F.) bound \cite{Breitenlohner:1982jf}.}
\beq\label{eq:twoquant}
-\frac{9}{4}<m^2<-\frac{5}{4}\ ,
\eeq
one obtains a sensible solution using either root of
(\ref{eq:MassDelta}) (for a discussion, see
e.g. \cite{Balasubramanian:1998sn}). In the window of two
quantizations, there are two field theories with operators of
different scaling dimensions, corresponding to the choice of boundary
conditions.

To construct a holographic superfluid one must search for solutions to
the equations of motion obtained from (\ref{eq:Action}) such that the
charged operator has a nonzero expectation value even after its
external source is removed.  Gravitationally this means that one is
searching for a black hole solution with scalar hair in asymptotically
AdS space.  In Minkowski space such hairy black holes do not exist,
but in \cite{Gubser:2008px} it was noted that the negative
cosmological constant may stabilize hair outside a black hole.

Throughout this work we will work in the so-called probe limit,
($\frac{\kappa_4^2}{q^2}$ is small), so that the backreaction to gravity
can be ignored. The gravitational solution involves a planar AdS-Schwarzschild metric,
\beq
ds^2=\frac{1}{z^2}(-f(z)dt^2+f(z)^{-1}dz^2+dr^2+r^2d\theta^2),
\eeq
where $f(z)=1-z^3$. In the above, temperature has
been absorbed in a rescaling to dimensionless coordinates.
After the rescaling, the only free parameter is the dimensionless
ratio $\mu/T$. Changing this ratio may be thought of as changing the
temperature (chemical potential) and keeping the chemical potential
(temperature) fixed. In this background, charged scalar hair may then
emerge depending on the temperature.

When studying homogeneous states it is possible to go beyond the probe
approximation and include back reaction from the bulk scalar and gauge
fields to the black hole metric \cite{Horowitz:2009ij}.  In doing so,
one verifies that working with an uncharged black hole is a good
approximation when the scalar field's charge is large
($\frac{\kappa_4}{q}\sim .01$) and mass is small.

In order to find the dual field theory operator expectation values we
need to solve the classical equations of motion in AdS space to obtain
the on shell fields.  In the probe approximation, the equations of
motion become
\begin{align}
0 &= \frac{1}{\sqrt{-g}}\partial_\mu(\sqrt{-g} g^{\mu\nu}\partial_\nu R)+m^2 R - R(\partial_\mu\chi-A_\mu)^2\label{eq:ELeq1}
\\
0 &= \frac{1}{\sqrt{-g}}\partial_\mu(\sqrt{-g} F^{\mu\nu}) - R^2(A^\nu-\partial^\nu\chi)\label{eq:ELeq2}
\\
0&=\partial_{\mu}(\sqrt{-g}R^2g^{\mu\nu}(\partial_{\nu}\chi-A_{\nu})),\label{eq:ELeq3}
\end{align}
where we have defined the real valued fields $R$ and $\chi$ according to the relation $\Psi=\frac{1}{\sqrt{2}}Re^{i\chi}$.

\begin{figure}[h]
\begin{center}
\includegraphics[scale=1]{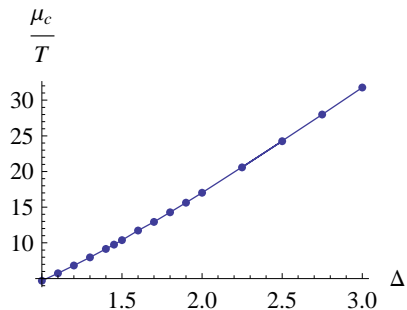}
\caption{\label{fig:MuCVersusDelta} The critical value of the chemical
  potential as a function of the charged operator's scaling dimension,
  while keeping the temperature fixed. $\Delta_-\ (\Delta_+)$ is to
  the left (to the right) of 1.5 on the horizontal axis.}
\end{center}
\end{figure}
The scalar fields are unstable to spontaneous symmetry breaking as one
raises the chemical potential $\mu$, so that above the critical value
$\mu_c$ the boundary theory enters the superfluid phase.
As we vary the mass of the bulk scalar field
over the range (\ref{eq:twoquant}), we cover the
following range of scaling dimensions:
$
 \frac{1}{2} < \Delta_- \leq \frac{3}{2}  \ ;
\   \frac{3}{2} \leq \Delta_+ < \frac{5}{2} \ .
$
Higher masses then correspond to the range $5/2 \leq \Delta_+$.

One effect of the varying scaling dimension is a change in the
critical value $\mu_c$ of the chemical potential where the superfluid
transition happens.  In Fig. \ref{fig:MuCVersusDelta} we have plotted
$\mu_c$ as a function of $\Delta$.  We see that operators of larger
scaling dimension have larger (smaller) values of $\mu_c$ ($T_c$).

\section{Solitons as a way to study the properties of the superfluid}

The last section described a set of holographic superfluids which may
be obtained by varying the scaling dimension and chemical potential.
There are many quantities that one could study, for example the
chemical potential dependence of the superfluid order parameter near
the critical temperature first found in \cite{Horowitz:2008bn}
\beq
\langle \mathcal{O}_\Delta\rangle \sim \sqrt{\frac{\mu}{\mu_c}-1}.
\eeq
A key question is what does one learn about the superfluids from such
computations. For example, the last scaling is presumably a mean field
result. For instance, one would like to know how to characterize the
degrees of freedom that are the constitutents of the superfluid (since
this is far from obvious in the holographic description).

There are at least three routes that one might consider.  The obvious
one is to come up with a detailed top-down model, with complete
control over the boundary theory at the microscopic level, as in the
original case of $D3$-branes and $N=4$ super-Yang-Mills theory (for
work in this direction, see
\cite{Gubser:2009qm,Gauntlett:2009dn,Gauntlett:2009bh}). However, most
models are bottom-up type, where the fundamental degrees of freedom
and dynamics of the boundary theory are unknown. That leaves one two
other strategies to consider.

The first is to consider linear response theory.  It is
straightforward to study the linear response of fields already
involved in the gravity solution, but one is also interested in the
way other fields, such as fermions, might respond to the superfluid.
This approach is obstructed because a bottom-up approach generally
lacks a stringy embedding which would dictate the allowed fermions.

The second, especially clean way to study holographic superfluids is
to study kinks and vortex solutions which asymptote to the homogeneous
ground states described in Section \ref{sec:Background}. The key
reason is that it is known that kinks and vortices may shed light on
the short distance features, even at the mean field level.
Essentially this is due to the fact that the core region must
interpolate all the way to the symmetric phase and hence the soliton
must know about physics of all length scales. Because one expects mean
field theory to still be sufficient, it is enough to look
for inhomogeneous solutions to classical gravity.  An advantage of
spatially dependent solutions is that they are inherent to the system --
one does not need to turn on any external fields to excite it.

In the rest of this Section we will discuss how to construct solitonic
solutions to the gravitational equations of motion.  Then we will
discuss several features of the solitons found in
\cite{Keranen:2009vi,Keranen:2009ss,Keranen:2009re} and an analogy
with BEC-BCS, which is useful both in organizing the holographic
results as well as suggesting further tests one might perform.

\subsection{Method\label{sec:Method}}

The differential equations obtained from (\ref{eq:Action}) are a set
of coupled nonlinear partial differential equations which are easiest
approached numerically.  The basic strategy will be to exploit the
fact that the equations are elliptic outside the horizon.  This will
allow us to construct an auxiliary heat equation which we solve
numerically.\footnote{It is instructive to remember how one might
  solve the Poisson equation \beq\label{eq:Poisson} \nabla^2\phi =
  0,\eeq by studying a heat equation \beq
  (\partial_\tau+\nabla^2)\phi(\tau,x).\eeq If we make any reasonable
  initial condition on the heat flow, at late times we flow to the
  "nearest" ground state satisfying (\ref{eq:Poisson}).  In this case
  nearest means in the same topological class.}

In more detail, as discussed in \cite{Keranen:2009re}, when we work in
the $A_z=0$ gauge and use the cylindrical symmetry the equations of motion
may be brought to the following form for vortex configurations,
\begin{align}
0&= f\partial_z^2\tilde{R}+(\partial_zf+(2\Delta_{-}-2)\frac{f}{z})\partial_z\tilde{R}-\Delta_{-}^2
z\tilde{R}+
\frac{1}{r}\partial_r(r\partial_r \tilde{R})\nonumber
\\
&-\tilde{R}(-\frac{1}{f}A_t^2+\frac{(A_\theta-n)^2}{r^2})\nonumber
\\
0 &= f\partial^2_z A_t+\frac{1}{r}\partial_r(r\partial_r
A_t)-z^{2\Delta_{-}}\tilde{R}^2A_t\nonumber
\\
0&= \partial_z( f\partial_z
A_\theta)+r\partial_r(\frac{1}{r}\partial_r A_\theta)
- z^{2\Delta_{-}}\tilde{R}^2(A_\theta-n),\label{eq:eqs}
\end{align}
where we have defined $\tilde{R}=z^{-\Delta_{-}}R$.  A similar set of
equations may be obtained for kink solutions as described in
\cite{Keranen:2009ss} for the case $m^2=-2$.

For both vortex and kink solutions we impose regularity conditions at
the horizon.  This is the same condition that was used to find the
homogeneous solution, and is necessary to obtain a steady state
solution. In the asymptotically AdS region we will impose a uniform
chemical potential as well as a vanishing scalar non-normalizable
mode.  The vortex solution also involves the $\theta$ component of the
gauge field, which we take to vanish at the AdS boundary.  This
corresponds to having no external driving vorticity to source the
vortex.

The basic strategy for finding solutions is to make an initial guess
at a solution (that respects the boundary conditions) and then let
this guess flow according to an auxiliary heat equation.  By waiting
long enough in auxiliary time, we arrive at a solution of the
equations of motion (\ref{eq:eqs}).  In reality all this is done on a
lattice and one stops the simulation when the evolution is suitably
slow. For details of the algorithm as well as error analysis see
\cite{Keranen:2009vi}.  Typical kink and vortex solutions are shown in
Figs. \ref{fig:KinkProfile} and \ref{fig:VortexCondensate}.

With this method we find that we are able to obtain good numerical
solutions for values of $T/T_c>.5$ (or $\mu/\mu_c<2$). See
\cite{Keranen:2009ss} for a detailed discussion of the numerics. The
numerical computations were performed with MATHEMATICA and some simple
C-programming using desktop computers.


\subsubsection{Solutions}
\paragraph{Kinks:}
\begin{figure}[h]
\begin{center}
\includegraphics[scale=1.2]{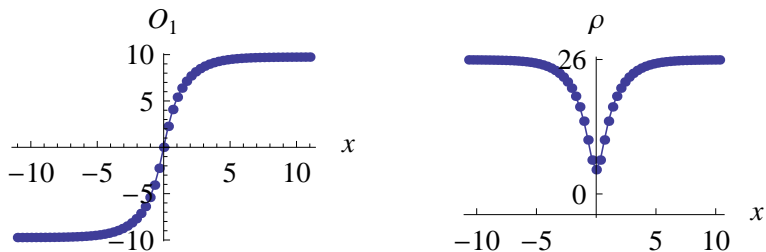}
\caption{\label{fig:KinkProfile} A typical kink soliton profile as
  computed using holography.  On the left is the condensate's profile
  plotted in the transverse coordinate.  On the right is the density
  profile in the same transverse direction.  These curves were
  obtained using a $\Delta=1$ condensate.}
\end{center}
\end{figure}

\begin{figure}[h]
\begin{center}
\includegraphics[scale=1.1]{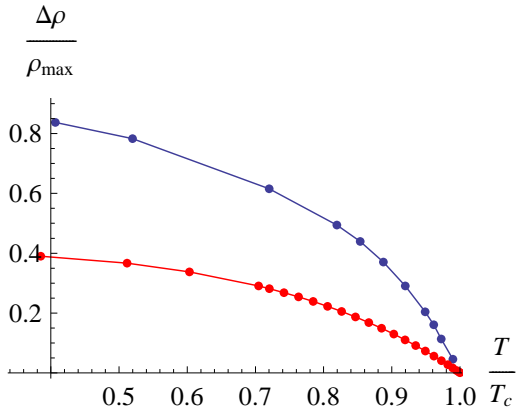}
\caption{\label{fig:OldKinkDepFrac} The core charge depletion fraction
  for kink solitons at $m^2=-2$ for varying $T/T_c$.  $\Delta = 1$ is
  indicated by the blue curve (top) and $\Delta = 2$ is indicated by
  the red curve (bottom).}
\end{center}
\end{figure}
In Fig. \ref{fig:KinkProfile} we have plotted a typical kink solution.
We see that the condensate passes through zero, as required by
topology and continuity.  We also see that the (number) density does
not completely vanish in the core of the soliton.
A nice feature of the holographic model is that it gives an effective theory for
the condensate {\em and} thermal fluctuations, so one can easily study the effects
of varying temperature.
In order to disentangle the thermal fluctuations, we reproduce
Fig. \ref{fig:OldKinkDepFrac} from \cite{Keranen:2009ss} displaying
the charge density's core depletion as a function of the temperature.
This graph indicates that the two superfluids have very different
behavior as one lowers the temperature.  Specifically, one finds that
the $\Delta = 2$ superfluid has an excess of charge density lying in
its solitons' cores in addition to the condensate's contribution.
This latter feature is totally different from what one would obtain
from a simple Gross-Pitaevskii picture for a purely bosonic condensate
at zero temperature, where the density goes to zero at the core.

\paragraph{Vortices:}
\begin{figure}[h]
\begin{center}
\includegraphics[scale=1.1 ]{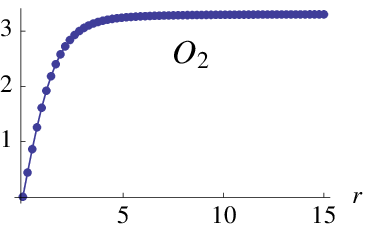}
\includegraphics[scale=1.1]{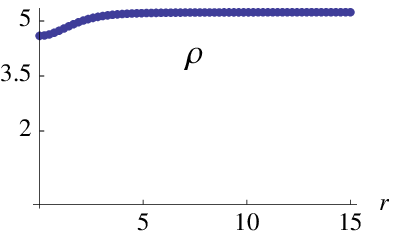}
\includegraphics[scale=1.1 ]{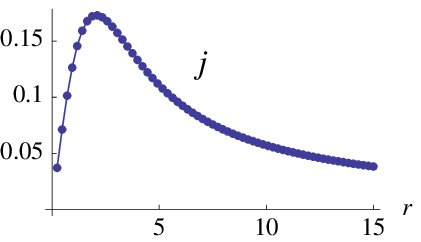}
\caption{\label{fig:VortexCondensate} A typical vortex soliton profile
  as computed using holography.  On the left is the condensate's
  radial profile.  In the middle is the profile of the charge density.
  On the right is the current density's radial profile.  These curves
  were obtained using a $\Delta=2$ condensate.}
\end{center}
\end{figure}
In Fig. \ref{fig:VortexCondensate} we have shown a typical vortex
profile.  Again, we find that the condensate vanishes in the vortex's
core as is expected.  We also find that the charge density does not
typically vanish in the core.  Finally, as discussed in
\cite{Keranen:2009re} we may identify the superfluid current from the
angular component of the bulk gauge field.  The superfluid current
rises from zero at infinity until the critical velocity is surpassed
and then falls to zero because the core is in the normal phase.

\begin{figure}[h]
\begin{center}
\includegraphics[scale=1.15]{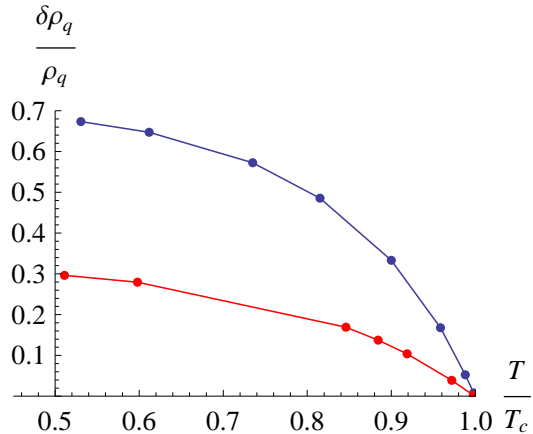}
\caption{\label{fig:OldVortDepFrac} The core charge depletion fraction
  for vortex solitons at $m^2=-2$ for varying $T/T_c$.  $\Delta = 1$
  is indicated by the blue curve (top) and $\Delta = 2$ is indicated
  by the red curve (bottom).}
\end{center}
\end{figure}

As for the kink solutions, we can again easily study the effect of varying temperature.
We have reproduced
Fig. \ref{fig:OldVortDepFrac} from \cite{Keranen:2009re} showing the
temperature dependence of the core charge depletion fraction.  This
figure also clearly shows that the $\Delta=2$ fluid has an "excess"
charge density in its core as compared to to the $\Delta=1$-fluid
In the next Section we will remind the reader of
what happens in the BEC-BCS crossover.  This will serve as a useful
guide to help interpret the core features of holographic solitons.

\subsection{BEC-BCS analogy\label{sec:Crossover}}
In the last Section we saw that, when $m^2=-2$, solitons have very
different core features when one changes the scaling dimension of the
condensing operator.  In this Section we wish to use the BEC-BCS
crossover as a benchmark system to help us organize our holographic
results.  This will help us both to interpret the features of the
holographic solitons as well as suggest further questions.

For the non-relativistic BEC-BCS crossover at zero temperature,
\cite{antezzaetal,Randerias} have studied the behavior of kink and
vortex solitons.  In both papers, it was found that soliton cores in
fermionic superfluids have non-vanishing number densities of cold
atoms, even though the condensates vanish.  This is very different
than BEC solitons, where the atomic number density vanishes (at zero
temperature) when the condensate vanishes.  In these systems soliton
cores are able to reveal the microscopic structure to the superfluid,
even at the mean field level.  For fermionic superfluids, as one
approaches the cores, there are additional states available in the
form of fermion excitations because the core is in the normal phase.
On the other hand, bosonic superfluid solitons have vanishing core
atomic number densities because there are no additional non-condensate
states available at zero temperature. In those works, it was also
found that the core structures smoothly interpolated between the BEC
and BCS limits. At finite temperature, even in the BEC superfluid, we
expect there will be a small contribution to the number density from
the thermal cloud (non-condensate normal fluid component) which can
show up in the soliton's core.

In both \cite{antezzaetal,Randerias}, it was argued that the solitonic
cores also were able to reveal the character of the additional states
carried by a fermionic superfluid's core.  Specifically, in
\cite{antezzaetal}, it was found that the solitons display Friedel
oscillations in their cores.  In \cite{Randerias} it was found that
the oscillations in vortices were finite size effects, not Friedel
oscillations.  On the other hand, \cite{Randerias} found that the
vortex cores knew about $k_f$ in the fact that the vortices had
different length scales in the cores and tails.

Comparing the holographic superfluids to the crossover systems
suggests that we should interpret excess core charge density in a
soliton as a signal that there are additional non-condensate states
residing in the core.  Therefore, we might expect to see that one can
interpolate between "empty" and "full" cores as one changes the
gravitational mass parameter hence changing the scaling dimension
\cite{Nishida:2006a,Nishida:2006b} .  In addition, we may also look
and see if holographic solitons identify the fermionic/bosonic
character of the "excess" states in the soliton's core.  To do this
there are at least two ways to proceed.  First one may try to identify
whether vortices display single or multiple length scales in their
cores and tails.  Second, one could hope to identify any Friedel
oscillations.

\subsection{Varying the Scaling Dimension}
\begin{figure}[h]
\begin{center}
\includegraphics[scale=1]{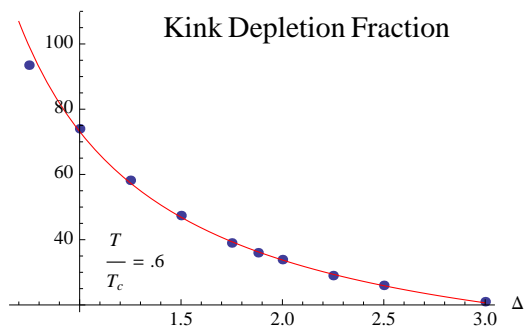}
\caption{\label{fig:KinkDepFrac} The charge density depletion fraction
  ($\times 100$) for kink solitons as a function of the condensate's
  scaling dimension at $T/T_c=.6$.  The curve is a best fit obtained
  for scaling dimensions above the B.F. bound ($\Delta = 3/2$). }
\end{center}
\end{figure}

In Subsection \ref{sec:Crossover} it was pointed out that comparing to
the crossover indicates that one should expect that the charge density
depletion fraction in a soliton's core to interpolate between $0$ and
$100\%$ as one varies the scaling dimension of the condensing operator
(at zero temperature).  In principle, it is simple to repeat the
analysis in Subsection \ref{sec:Method} when $m^2=-2$ now allowing for
more general scaling dimensions. In practice, one cannot work at $0$
temperature for two reasons.  First, we have assumed a probe
approximation which is only valid for temperatures moderately below to
$T_c$ ($\mu$ above $\mu_c$).  Secondly, even if we went beyond the
probe approximation, it is computationally too expensive to obtain
solutions for arbitrarily low temperatures.

In order to proceed we note that in Fig. \ref{fig:OldKinkDepFrac} and
Fig. \ref{fig:OldVortDepFrac} the depletion fraction seems to be
saturating near $T/T_c\sim .5$ independent of the scaling dimension.
Therefore, as an approximation to the depletion fraction at low
temperature we will study solitons at $T/T_c = .6$ as we vary
$\Delta$.  Finally, for the sake of brevity we will focus on the kink
solitons.  It is a straightforward exercise to show that vortices also
display the same features as one varies $\Delta$.

In Fig. \ref{fig:KinkDepFrac} we see the smoothly varying charge
density depletion fraction in the core of a kink soliton at fixed
$T/T_c=.6$.  A numerical fit of the form
\beq
\frac{\delta\rho}{\rho_{max}} = A+\frac{B}{\Delta}
\eeq
was made for the points with scaling dimension greater than $\Delta =
3/2$.  The best fit values are $(A,B)=(-.06,.8)$. The small variation
in the graph is primarily due to the small uncertainty in the value of
$T/T_c$.  We note that there is a visible change in the behavior of
the depletion fraction as one lowers $\Delta$, presumably saturating
close to $100\%$ at $\Delta \sim 1/2$, although these values are
beyond what we can currently simulate.

\section{Characterizing the core states\label{sec:LinearResponse}}

\subsection{Multiple length scales}
Having seen that one can indeed control a soliton's core charge
depletion fraction by varying the condensing operator's scaling
dimension, and hence, the number of states available in the soliton's
core, we would also like to see to what extent one may try to
characterize these states.  The simplest way one might do this is to
obtain characteristic length scales for the condensate's profile in the
core and in the asymptotic tails.  In \cite{Randerias} it was found
that bosonic superfluids essentially have one length scale, while
fermionic superfluids lead to distinct length scales in the two
regimes.
\begin{figure}[h]
\begin{center}
\includegraphics[scale=1.1]{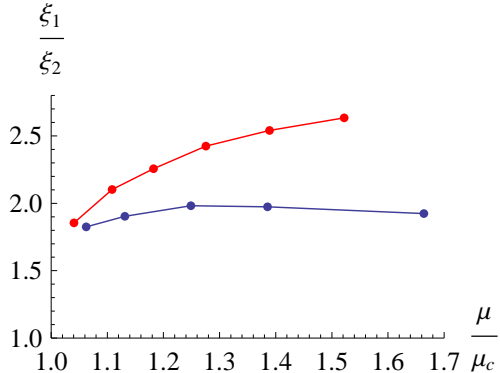}
\caption{\label{fig:XiRatio} Here we have reproduced the graph from
  \cite{Keranen:2009re} plotting the ratio of length scales of
  variation as determined from the condensates profile for vortex
  solutions.  The blue curve (bottom) indicates a $\Delta=1$
  condensate and the red curve (top) indicates a $\Delta=2$
  condensate.}
\end{center}
\end{figure}
We may try to identify how dissimilar the soliton's length scales are
according to how their ratio varies as a function of $\mu$ for
different operator scaling dimensions.  Fig. \ref{fig:XiRatio} was
obtained in \cite{Keranen:2009re}, indicating that there are two
distinct length scales as one increases the scaling dimension.  It
would be very interesting to see what happens to the corresponding
graph for $\Delta\sim 1/2$, the lowest allowed value.  The comparison
with the crossover would predict that in this regime the corresponding
curve would be completely flat.  Unfortunately, we do not have
resources to say more about this value of $\Delta$.  In Subsection
\ref{sec:LinearResponse} we will discuss critical velocities obtained
from linear response theory.  Comparing these critical velocities to
the superfluid's sound modes will lead to a morally similar comparison
to Fig. \ref{fig:XiRatio}.

\subsection{Friedel Oscillations}

A final feature of solitons in the crossover was the possible
observation of Friedel oscillations. As discussed in
\cite{Keranen:2009ss,Keranen:2009re}, we find no evidence of Friedel
oscillations in the holographic solitons. One possible reason for the
absence of the oscillations, if they exist, is that they may simply be
obscured by a too large temperature.


\subsection{Critical velocity}\label{sec:CriticalVelocity}

Another way to probe the constituents of a superfluid is to study
superfluid flows. As one increases the velocity of the flow, it
becomes energetically more favorable for the superfluid to radiate
quasiparticles and go to the normal phase. The velocity at which this
happens is called the critical superfluid velocity. It is possible to
estimate the critical velocity using a simple kinematical argument due
to Landau. According to Landau's criterion, the critical superfluid
velocity is given by the formula
\beq
v_c=\min_{k}\frac{\epsilon(k)}{k},
\eeq
where $\epsilon(k)$ is the quasiparticle dispersion relation in the
superfluid and the minimum is taken over all the
quasiparticles. Strictly speaking the Landau criterion only gives an
upper bound for the critical velocity and indeed at finite temperature
one expects to see a smaller critical velocity than the Landau
criterion tells us \cite{Navez}. Still the Landau criterion can act as
a useful guide in estimating the critical velocities.

For a BCS superfluid, the critical velocity is set by the lightest
fermionic excitations which leads to an estimate
\beq
v_c^{BCS}\approx\frac{\omega_{gap}}{k_f},
\eeq
where $k_f$ is the Fermi momentum and $\omega_{gap}$ is the energy gap
for fermions in the superfluid phase. On the other hand on a BEC
superfluid the critical velocity is set by the sound modes with lowest
sound velocity $v_s$ so that
\beq
v_c^{BEC}\approx v_s.
\eeq
Unfortunately in the holographic superfluid model we do not have a
direct access to possible fermions in the system, without adding extra
fields to the bulk. Even though one can access some fermionic features
of the system by adding probe fermions to the bulk, the connection of
these fermions to the ones possibly comprising the condensate is not
very clear. Thus, we are unable to calculate the fermion dispersion
relations required to get the "fermionic" Landau criterion.

Instead we can calculate sound velocities and compare them to the
critical velocity obtained from the vortex solutions. A study of the
Landau criterion in the BEC-BCS crossover can be found in
\cite{Randeria2}.

According to \cite{Herzog:2009md} the second sound has the lowest
sound velocity, at least in the part of parameter space studied
there. As shown in \cite{Herzog:2008he} the second sound can be
calculated from thermodynamic quantities as
\beq
v_2^2=-\frac{\Big(\frac{\partial^2P}{\partial\xi^2}\Big)_{\mu,T}}
{ \Big(\frac{\partial^2P}{\partial\mu^2}\Big)_{\xi,T}}.
\eeq
Here $\xi=\nabla\chi$, where $\chi$ is the phase of the condensate and
the thermodynamic derivatives are evaluated at $\xi=0$. More details
on the calculation may be found in \cite{Herzog:2008he}.


The critical superfluid velocity may be obtained from the vortices as
is described in \cite{Keranen:2009re}. One simply looks for the radial
position $\rho_*$ inside the vortex where the condensate vanishes and
evaluates the superfluid velocity at that radius. This leads to
\beq
v_c=\frac{1}{\mu}\frac{n}{\rho_{*}}.
\eeq
A convenient way to determine $\rho_*$ is to identify it with the
position of peak current \cite{Randerias,Keranen:2009re}.

\begin{figure}[h]
\begin{center}
\includegraphics[scale=1.1]{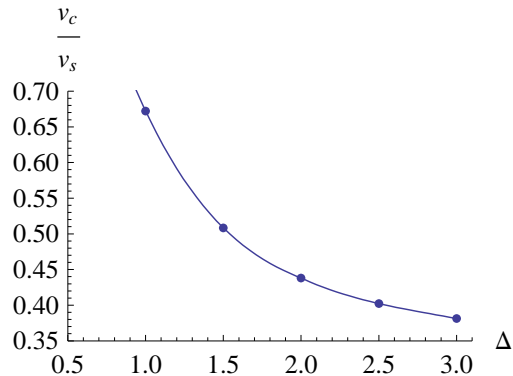}
\caption{\label{fig:velratio} The ratio of the critical superfluid
  velocity and the second sound velocity as a function of the
  condensate scaling dimension. The curve in the figure is an
  interpolating function while the dots are the real data points.}
\end{center}
\end{figure}

The ratio of the vortex critical velocity to the sound velocity is
plotted in Fig. \ref{fig:velratio} for $T/T_c\approx 0.7$.
We find that the ratio of the two velocities seems to be approaching 1
as the scaling dimension of the condensing operator is lowered, and is
notably smaller than 1 as the dimension is larger. In a BEC like
superfluid one would expect a sound mode to determine the critical
velocity. Recalling that in our crossover analog the small scaling
dimensions correspond to the supposed BEC region, we indeed find a
consistent picture with the analog. In a BCS like superfluid one
expects the critical velocity to be set by a fermionic excitation
rather than a sound mode. Indeed it seems that in the supposed BCS
region the critical velocity is likely set by something else than a
sound mode.

\section{Discussion}

The possibility to apply holographic techniques to low dimensional
systems is an exciting recent development in string theory.  However,
because the set of quantities which are readily computable are
typically quite different, it can be difficult to develop necessary
model building intuitions.  For this purpose it is often useful to
compare holographic constructions to known real world systems which
display many features expected of holographic systems.  In this
article we have focused on using the BEC-BCS crossover as a guide to
features observed in holographically constructed superfluids.  This is
a crossover from a system of fundamental bosons to fundamental
fermions, both of which may display superfluidity.  Most relevant for
holography, this is a system which is strongly interacting and may be
studied by moving along a family of fixed points \cite{Sachdev}.

We began by constructing kink and vortex solitons in holographic
superfluids obtained by condensing operators of scaling dimensions
$\Delta=1$ and $2$.  Surprisingly their cores display different
features, with the $\Delta=2$ soliton cores supporting much larger
charge densities.

Similar features occur in the BEC-BCS crossover, where there are
additional non-condensate states available away from BEC regime.  In
this setting the additional states are comprised of pre-formed bosons
and fermions.  Also, the solitons "know" about the fermions in the BCS
regime because the Fermi momenta controls the core size as well as any
Friedel oscillations.

Comparing with the crossover physics suggested that as we vary the
condensing operator's scaling dimension we should expect the core
features to smoothly interpolate as we move along the family of
CFT's. Indeed, this is precisely what was obtained.  As we increased
the scaling dimension, the amount of charge supported by soliton cores
monotonically increased.  Also consistent with this trend is the fact
that the difference between core and tail length scales of variation
increases with $\Delta$ \cite{Keranen:2009re}.  Finally, from the
solitons we find no signature of Friedel oscillations for $T/T_c>.5$.

In an effort to get a handle on the character of the excess charged
states available in soliton cores we also compared the critical
velocity obtained from vortex cores to the Landau critical velocity.
As one lowers $\Delta$ the vortex critical velocity approaches the
Landau critical velocity monotonically.  This is what one would expect
for a BEC superfluid.


The BEC-BCS analogy may also be used to explain results found in
\cite{Horowitz:2008bn} for conductivities calculated on the superfluid
phase. There it was found that there are delta function like spikes on
the frequency dependent conductivity. These spikes were interpreted as
due to bound states. The only bound states that may contribute to the
conductivity (current-current correlator) are those that have a
vanishing net charge. Thus, a natural candidate for the spikes on the
BEC-BCS picture is a bound state of a fermion-hole pair. Such pairs
should exist on the BCS side only. Indeed it was found in
\cite{Horowitz:2008bn} that the spikes appear on the "large" scaling
dimension side, and the binding energy calculated from the position of
the spikes seems to increase as the scaling dimension is lowered. This
is consistent with the crossover picture where the interactions
increase as one approaches unitarity from the BCS
side. Holographically, as one moves across the B.F. bound
($\Delta=3/2$) to lower scaling dimensions, at some point the spikes
disappear, which would seem to signal that the bound state
disappears. This fits well with the BEC-BCS analogy since on the BEC
side (the low scaling dimensions) the Fermi surface and light
fermionic excitations are expected to completely vanish. The crossover
analogy would naturally explain why the particle-hole bound state
would go away.

We have shown that the BEC-BCS may be a useful guide to holographic
superfluids, owing to their common strong coupling and conformal
features.  They also share common structures in their solitons.
However to make the crossover more than an interpretational guide it
would be more convincing if there was a direct signature of fermionic
features in holographic superfluids.  There are at least two places
one might hope to find such features, first if one could cool solitons
to low enough temperatures it might be possible to observe Friedel
oscillations.  This would be an unambiguous signal.  A second route
would be to try to have a better understanding of fermion probe
calculations, with the hope that they could reveal any fermionic
feature encoded in the bulk geometry.

\section{Acknowledgements}
V.K. and E.K-V. have been supported in part by the Academy of Finland
grant number 1127482.  E.K-V. and S.N. thank the Galilei Institute for
Theoretical Physics for their hospitality and the INFN for partial
support during the completion of this work. V.K. thanks the Nordic
Institute of Theoretical Physics (NORDITA) for their hospitality and support
while finishing this article. K.P.Y. thanks the Helsinki Institute of Physics for hospitality
while this work was in progress.

\end{document}